\documentclass[iop]{emulateapj} 

\usepackage{graphicx}
\usepackage{color}
\usepackage{kpfonts}
\usepackage{xspace}

\definecolor{dartmouthgreen}{rgb}{0.05, 0.5, 0.06}

\newcommand{\sS}[1]{\mbox{$\rm{}^{#1}$}}
\newcommand{\Ss}[1]{\mbox{$\rm{}_{#1}$}}
\newcommand{\um}{\mbox{$\mu$m}\xspace}

\begin{document}

\title{Miniature lightweight x-ray optics (MiXO) for surface
elemental composition mapping of asteroids and comets
}


\author{Jaesub Hong\sS{1*}, Suzanne Romaine\sS{2} and the MiXO team \\
              \sS{1} 
	      Harvard University,
              60 Garden Street, Cambridge, MA 02138, USA,
	      (jhong@cfa.harvard.edu) \\
	      \sS{2} 
	      Smithsonian Astrophysics Observatory,
              60 Garden Street, Cambridge, MA 02138, USA
	      (sromaine@cfa.harvard.edu) \\
	      \sS{*} Send requests to J. Hong.
}



\begin{abstract} 

\ \ \ The compositions of diverse planetary bodies are of fundamental
interest to planetary science, providing clues to the formation and
evolutionary history of the target bodies and the Solar system as a whole.
Utilizing the X-ray fluorescence unique to each atomic element, X-ray
imaging spectroscopy is a powerful diagnostic tool of the chemical and
mineralogical compositions of diverse planetary bodies.  Until now the
mass and volume of focusing X-ray optics have been too large for
resource-limited in-situ missions, so near-target X-ray observations
of planetary bodies have been limited to simple collimator-type X-ray
instruments.

\ \ \ We introduce a new Miniature lightweight Wolter-I focusing X-ray Optics
(MiXO) using metal-ceramic hybrid X-ray mirrors based on electroformed
nickel replication and plasma thermal spray processes.  MiXO can enable
compact, powerful imaging X-ray telescopes suitable for future planetary
missions. We illustrate the need for focusing X-ray optics in observing
relatively small planetary bodies such as asteroids and comet nuclei.
We present a few example configurations of MiXO telescopes and demonstrate
their superior performance in comparison to an alternative approach,
micro-pore optics, which is being employed for the first planetary
focusing X-ray telescope, the Mercury Imaging X-ray Spectrometer-T (MIXS-T)
onboard {\it Bepicolumbo}.  
X-ray imaging spectroscopy using MiXO will open
a large new discovery space in planetary science and will greatly enhance our
understanding of the nature and origin of diverse planetary bodies.

\end{abstract}

\section{Introduction}
\label{intro}

\begin{figure} \begin{center}
\includegraphics[width=3.2in,trim=0 0.9in 0 0.7in]{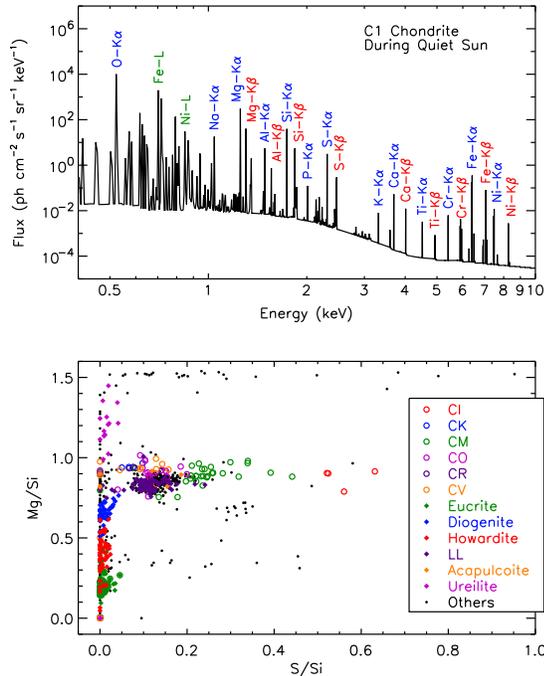} 
\caption{ (Top)
Simulated X-ray fluorescence spectrum of an asteroid of C1 chondrite composition
at 1 AU during the quiet sun state illustrating diverse elemental
composition. (Bottom)
Abundance ratios (Mg/Si vs. S/Si) as an identifier of a wide range of
meteorite specimen types \citep{Nittler04}. }
\label{f:spec}
\end{center}
\end{figure}

{\it ROSAT}, {\it the Chandra X-ray Observatory} and other X-ray
observatories have revealed that most planetary bodies
in our Solar System 
emit X-rays via various physical processes
including X-ray fluorescence (XRF) and charge exchange with plasma.
In XRF, the observed energies are unique
to the atomic levels of each element. This property of XRF enables a
powerful way to probe the surface distribution of elemental
composition in the emitting bodies.
Fig.~\ref{f:spec} shows a simulated X-ray spectrum from an asteroid of C1
chondrite composition located at 1 AU during a typical quiet sun state. X-ray
emission from the asteroid contains numerous XRF emission lines  tied
to various atomic elements
\citep{Nittler04}.  
Fig.~\ref{f:spec} also shows how the relative abundance of Mg/Si vs.~S/Si
can be used to identify matching meteorite specimen types.

While Earth-orbiting X-ray observatories are capable of high
angular resolution X-ray imaging (e.g.~sub-arcsec angular resolution
for {\it Chandra}), the large distance to relatively
X-ray faint rocky airless bodies
severely limits the photon statistics and often makes it difficult to
identify interesting surface features or substructures of X-ray emission. 
For instance, the observed X-ray flux of the Crab nebula, one of the
brightest X-ray sources in the sky is a few photons cm\sS{-2}
s\sS{-1} keV\sS{-1} at around 1 keV. This is comparable to the XRF flux
from an asteroid located at 1 AU from the Sun observed by a telescope
with a FoV of $\gtrsim$ 10 deg $\times$ 10 deg at the surface.  Due to
the Lambertian reflectance, the XRF flux observed by such a telescope
will be more or less the same regardless of the distance as long as
the FoV is fully occupied by the emitting surface.  On the other hand,
if the asteroid of $\sim$ 1 km size is observed at a distance of roughly
1000 km, only a tiny fraction of the XRF flux (10\sS{-5}) will reach
the telescope since the emission of the XRF is more or less
isotropic.

X-ray optics in the past are massive, so
X-ray instruments flown on
near-target missions in the past have foregone true imaging capabilities.
Instead, they had to employ simple collimator-type
instruments as in the X-ray spectrometers (XRS) 
on {\it NEAR-Shoemaker} \citep{Trombka00},
{\it Hayabusa} \citep{Okada06} or {\it Messenger} \citep{Nittler11}.  
This is largely because a large portion of the resources in 
a typical planetary mission
often has to be reserved for the spacecraft to make
possible the long journey to the target and the remaining resources are
shared by a collection of onboard instruments required to unveil the
mysteries of the targets.
Hence
the resources allowed for each instrument are limited, and 
conventional X-ray focusing optics used in the past are
simply too large and too heavy to be a part
of planetary missions.

Despite the lack of imaging capability, X-ray observations of the targets
have been proven to provide unique insights.
For instance, XRS observations of 433 Eros by {\it NEAR-Shoemaker}
revealed that the elemental composition of the asteroid from the X-ray
spectra is different from the asteroid classification acquired by
the optical and infrared spectra \citep{Trombka00}.
It is suspected that ongoing space
weathering alters the surface of the asteroids up to a few $\mu$m depth, which 
modifies the optical and infrared spectra, whereas the XRF probes
more than 20 $\mu$m below the surface and can reveal the true,
unweathered, composition \citep{Binzel10}.
In the case of Itokawa, definitive evidence of space
weathering on the asteroid was later found in the returned sample \citep[2014]{Noguchi11}.

\begin{figure} \begin{center}
\includegraphics[width=3.0in]{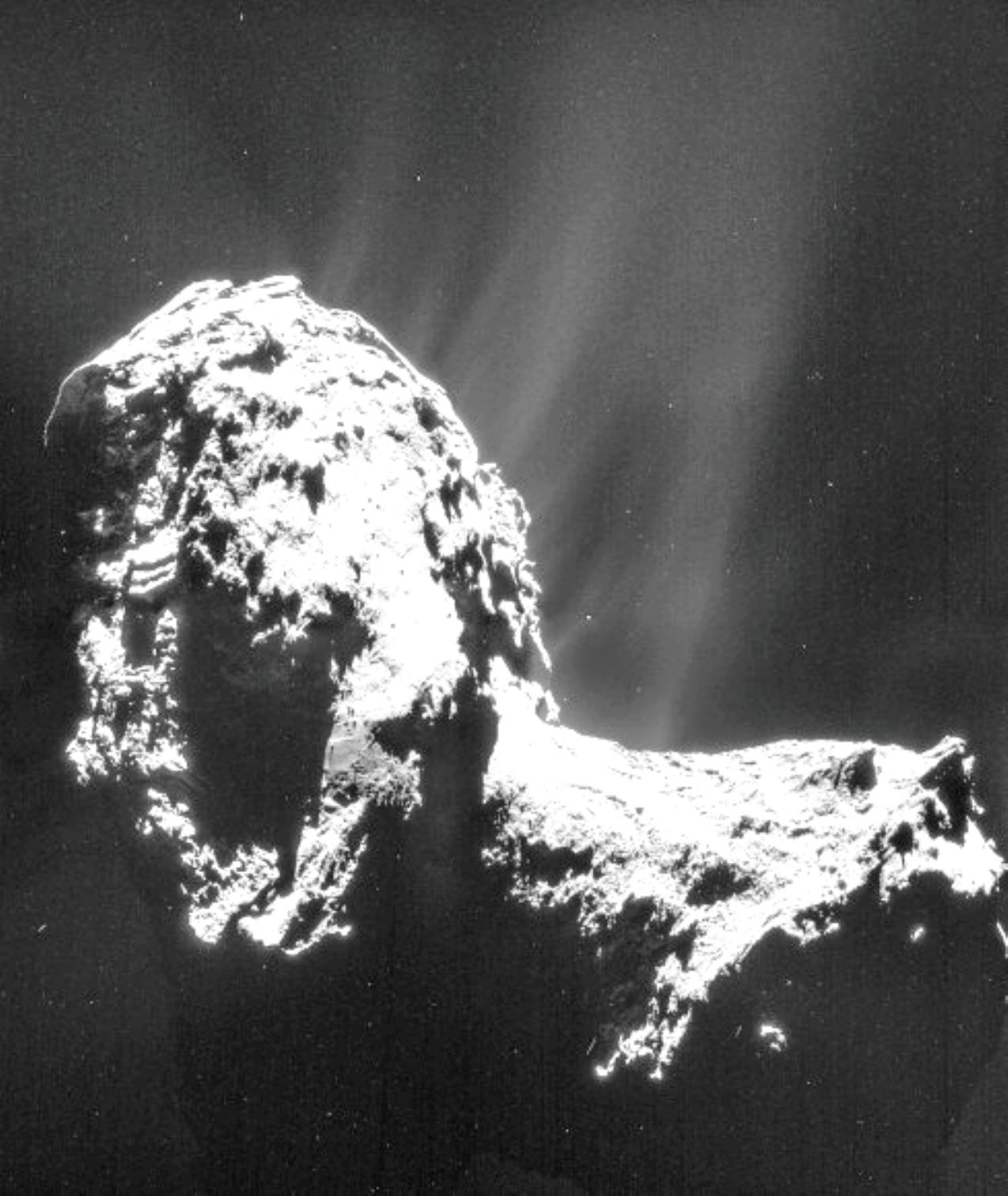}
\caption{
Optical image of Comet 67P/Churyumov-Gerasimenko taken
by NAVCAM on the Rosetta mission reveals the outgassing jets
(Courtesy of European Space Agency - ESA). 
}
\label{f:67P}
\end{center}
\end{figure}

From recent images of Comet 67P/ Churyumov-Gerasimenko taken by NAVCAM
($\sim$ 18\arcsec resolution)
on the Rosetta mission one can imagine exciting potential of future
X-ray observations of asteroids and comets when
X-ray spectroscopy is combined with high resolution imaging.  
The optical image of Comet 67P shows 
diverse terrain and volatile activity (Fig.~\ref{f:67P}).
In particular, the high ratio of deuterium in the water vapor of the
jets questions our understanding of the origin of the water on Earth
\citep{Altwegg15}.  While soft X-ray observations (0.1--15 keV) are
insensitive to isotope variation, X-ray imaging spectroscopy can
separate the elemental
composition of gaseous jets from the comet's surface as well as the
compositional variation of the surface due to impacts and volatile
activities in the past.  

\begin{figure*} \begin{center}
\includegraphics[width=6.2in]{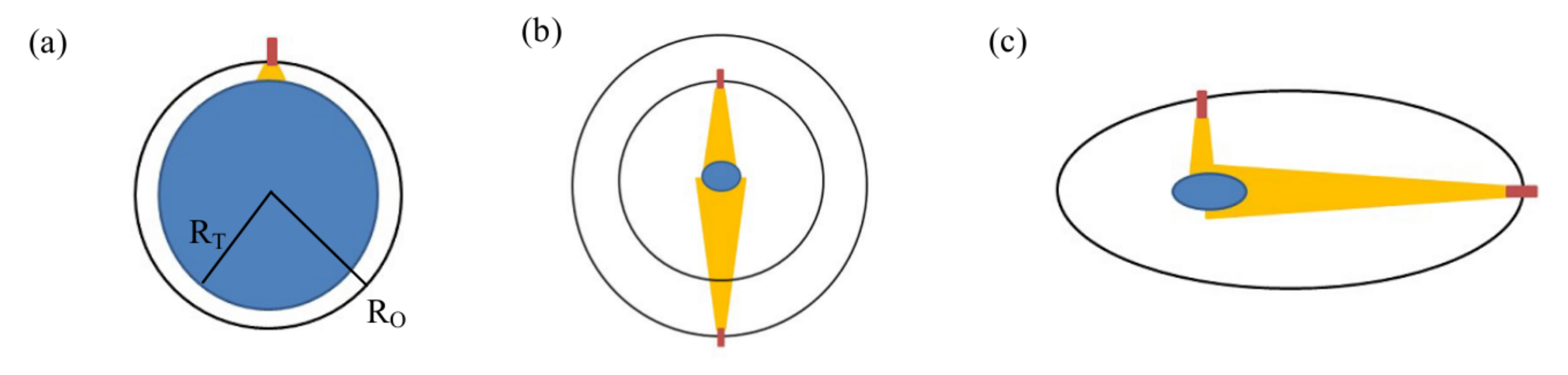} \\
\caption{(a) For large spherical planetary bodies (radius R\Ss{T} $>$
$\sim$1000 km), observation can be conducted at a close stable orbit
(R\Ss{O}) (i.e.  R\Ss{T} $\sim$ R\Ss{O}), where a simple collimator
can be sufficient for identifying large surface variation. (b) \& (c) For small targets
(R\Ss{T} $<$ $\sim$100 km), low gravity or irregular shape prevents
observation at a close steady orbit (i.e. R\Ss{T} $<<$ R\Ss{O}), and
the size of the imaging footprint on the surface can change
dramatically over time or even in a single orbit \citep{Hong14}. }
\label{f:orbit}
\end{center}
\end{figure*}

As the need for sample
retrieval of various planetary bodies rises for in-depth analysis
in attempt to unambiguously unveil their nature, 
origin and evolutionary history,
many future sample return planetary missions are being
developed.
The retrievable size of samples, however, is very small:
e.g.~sub-gram in Hayabusa 2 \citep{Tachibana13}, less than 1 kg
in OSIRIS-REx \citep{Lauretta12}. Therefore, 
X-ray imaging spectroscopy onboard future
near-target missions becomes an indispensable tool  in  acquiring a
global context of the atomic composition for the returned samples as
well as assisting the sample site selection during the mission.

In Sec 2, we review the major challenges encountered in near-target
X-ray observations using collimator instruments, and we illustrate how
focusing X-ray optics can overcome these challenges and enhance a chance of
success and new discoveries 
especially in observing relatively small bodies such as
asteroids and comet nuclei.  In Sec 3, we briefly review the advances of
Wolter-I X-ray optics, and introduce our new approach to build compact,
lightweight X-ray optics by merging Electroformed Nickel Replication (ENR)
and Plasma Thermal Spray (PTS) processes.  In Sec 4, we show example
MiXO configurations suitable for planetary missions and estimate the
performances in comparison to an alternative approach, micro-pore optics
(MPO).

\section{Advantages of focusing X-ray optics}
\label{s:optics}

X-ray observations in near-target missions in the past have been
challenging. Some of the challenges originate from incomplete prelaunch ground
calibration of the instruments and inadequate onboard calibration system
(usually due to tight assembly schedule and budget constraints). As a
result, the data analysis was complex and often led to somewhat
ambiguous interpretation.  On the other hand, a cause of many
challenges in the past planetary X-ray observation lies in the very
nature of non-focusing optics of collimator instruments as described below.

REgolith X-ray Imaging Spectrometer (REXIS) onboard OSIRIS-REx, which is
scheduled to launch in 2016 to collect samples from Bennu, was designed
to take into account lessons learned from the previous missions
\citep{Inamdar14}. For instance, REXIS will
be equipped with one of the most comprehensive onboard calibration systems in
space X-ray instruments: a set of \sS{55}Fe radioactive sources are
strategically mounted to monitor any potential drifts or changes
of spectral gain and resolution of every node of 4 X-ray CCDs
onboard.  In addition, the team plans to conduct a
series of extensive ground and flight calibrations to understand
the detector response and to track any variation 
during the observation.

REXIS utilizes coded-aperture imaging to identify $\sim$ 50 m scale
surface variation of the elemental composition of Bennu.  The addition
of imaging capability
is the first of its kind (coded-aperture imaging)
in planetary X-ray observations (see below about the {\it Bepicolumbo} mission).
Coded-aperture imaging is basically shadowgram imaging using a mask 
coded with a pattern of open and opaque elements
in front of the detector. It is a novel technique to enable imaging 
without focusing optics, and thus allows a relatively compact, simple
telescope design (REXIS fits in an envelope of $\sim$ 10 cm $\times$
10 cm $\times$ 20 cm).  However, coded-aperture imaging
still operates under the basic principle of collimator instruments, 
facing the same challenges. Thus its imaging sensitivity is relatively low
compared to focusing instruments and it has been mainly employed
to identify point sources in astrophysics in the past.

When observing large planetary bodies such as the Moon, a simple
collimator based X-ray instrument can efficiently collect the X-rays
from the surface and identify spatial variations as the instrument
orbits around the target. The large gravity and the
spherical geometry of the target with a small eccentricity allows stable
circular orbits very close to the surface (Fig.~\ref{f:orbit}a). In addition, the field of
view (FoV) of the instrument is fully occupied by the target,
so that no external background can contaminate the data.

\begin{figure} \begin{center}
\includegraphics[width=3.2in]{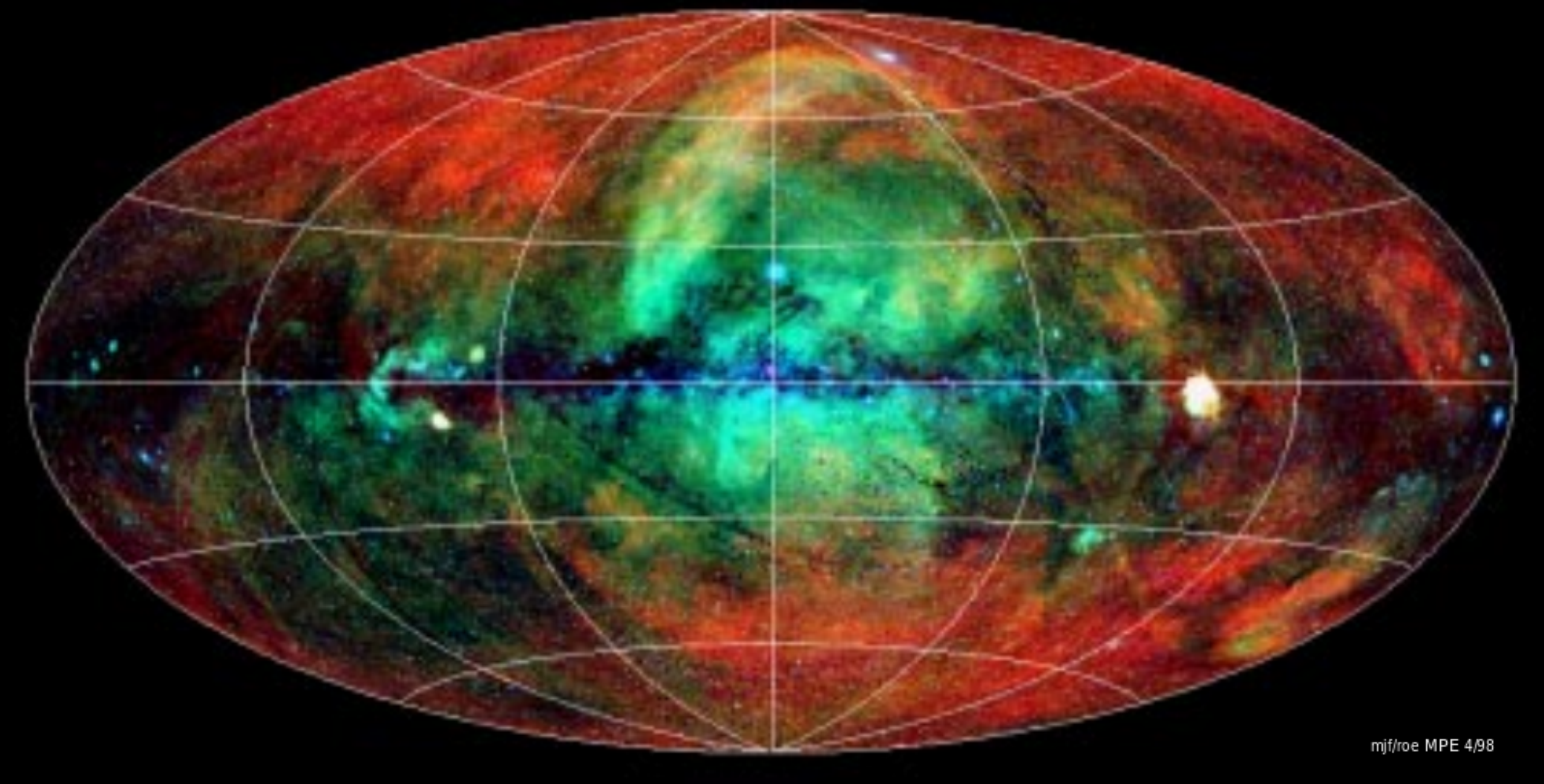} \\
\caption{All sky soft X-ray background image in Galactic coordinates
by ROSAT/PSPC 
after removing bright point sources (red: 0.1-0.4
keV, green: 0.5-0.9 keV, blue: 0.9-2.0 keV,
Courtesy of Max-Planck-Institut for extraterrestrische Physik).
}
\label{f:sky}
\end{center}
\end{figure}

In the case of relatively small planetary bodies such as asteroids and
comet nuclei, the small gravity and the irregular
shape of the target often limit stable orbital configurations near the target
(Fig.~\ref{f:orbit}b \& c).  In order to collect faint X-ray emission
from the target, the FoV of the instrument is usually designed to fill
the target, and thus the collimator type instrument cannot efficiently
identify the surface feature of small scales.  At the same time, the 
exposure to the background sky in the FoV should be limited because
the X-ray sky is not dark, but it shines with many diffuse and point
X-ray sources, often
brighter than the target itself. For instance, 
our Galactic plane glows brightly in X-ray, which 
is known as Galactic ridge X-ray emission from the early
days of the X-ray astronomy.
The sky at high Galactic latitude
is also filled with X-ray emission from unresolved Active Galactic Nuclei
(Fig.~\ref{f:sky}). These components contribute to bright Cosmic X-ray background
(CXB), which can often dominate X-ray emission from the target 
in the collimator-type instruments when their FoV is exposed to the
background sky. The continuum component of the CXB is
about 10 ph cm\sS{-2} s\sS{-1} sr\sS{-1} keV\sS{-1} at 1 keV,
which is about 100 -- 1000 times brighter than the example in
Fig.~\ref{f:spec}.

X-ray instruments in space also experience additional internal
X-ray background which occurs due to interactions between cosmic-ray
particles and the instruments.  The internal background can contain XRF
lines originated from the instrument, which can complicate the elemental
identification 
of the target.  The sensitivity of a typical collimator
instrument is proportional to a square root of the detector area
(instead of being linearly proportional to the detector area) because
the internal background increases as the detector area increases.
On the other hand, focusing telescopes allows a small focal plane with
a large effective area, so that the internal background can be 10-100
fold smaller than the collimator instruments of similar collecting area.

\begin{figure} \begin{center}
\includegraphics[width=3.2in]{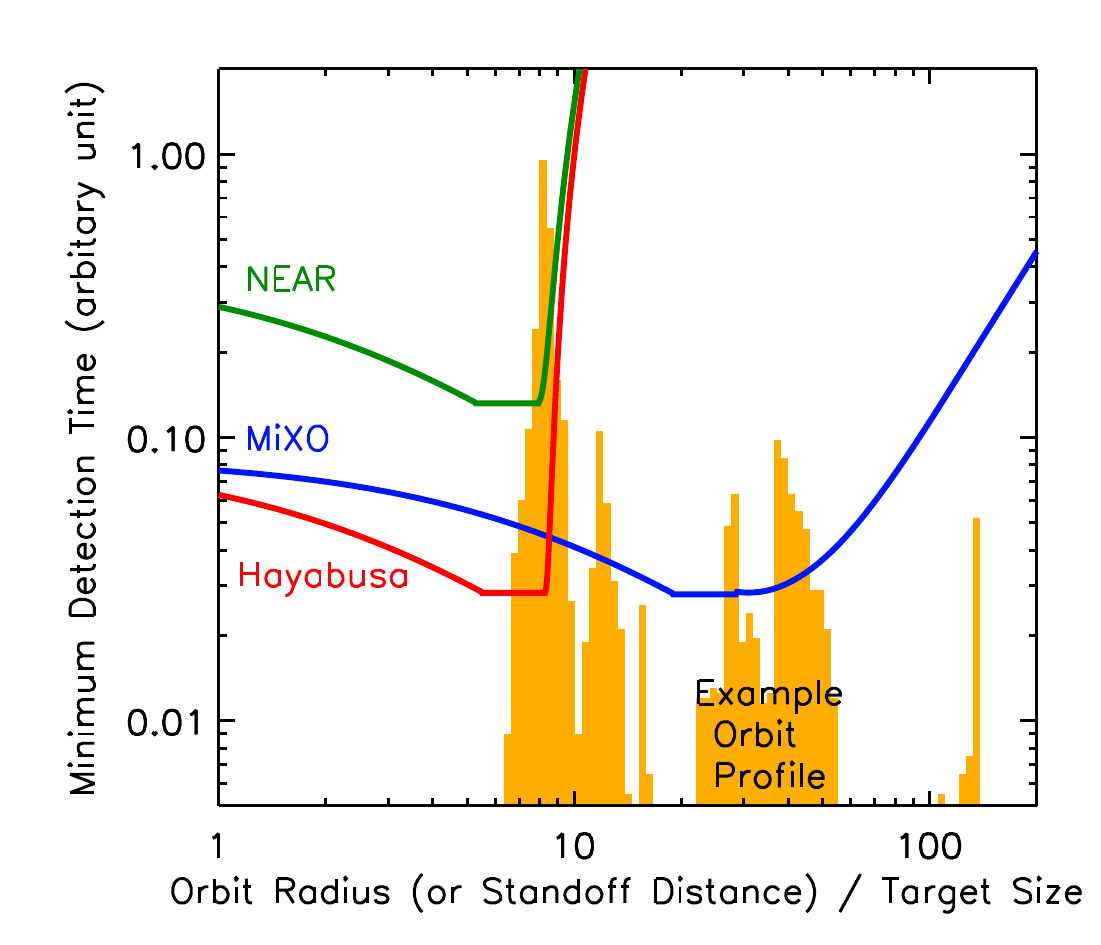} \\
\caption{Comparison of minimum detection time (global, non-imaging)
of XRF in 0.5--2 keV (depending on the optical/IR filtering scheme, mirror
reflectivity and other optics onfiguration) 
as a function of distance to the target for NEAR-Shoemaker
(green),Hayabusa (red) and MiXO (blue) with planned orbital profiles
(OSIRIS-REx, yellow).  The orbit radius and standoff distance are
rescaled to the target size. Collimator instruments become useless
beyond a certain distance due to high cosmic X-ray background, whereas
focusing optics can extend the useful observing condition to a much
wider range of orbit profiles, allowing a flexible mission design.
The minimum detection time is the total exposure that
is required to acquire the signal-to-noise ratio above a certain limit
(e.g.~5$\sigma$). 
}
\label{f:time}
\end{center}
\end{figure}

Due to high CXB (and the 
large internal background), observations with collimator instruments
are susceptible to changes in observing configurations
(e.g.~changes in pointing directions or observing distances),
which can severely narrow the observing windows.  Focusing telescopes,
on the other hand, can start meaningful observations much farther out
(e.g. during early phases of approach and debris scouting) by simply resolving
and removing most of both external and internal backgrounds. 

Fig.~\ref{f:time} compares relative minimum
exposures for XRF detection from an asteroid by well calibrated XRS of
{\it NEAR-Shoemaker} and {\it Hayabusa}
with an example MiXO telescope (MiXO-70B, see Sec.~4) as
a function of standoff distance or orbit radius (R\Ss{O}) 
normalized to the target size
(radius, R\Ss{T}).  It also shows a scaled example 
orbital/approach
profile from OSIRIS-REx (the orange histogram).
In-situ or near-target
experiments studying small asteroids (R\Ss{O} $\lesssim$ 100 km) often
cannot stay in a stable orbit at close proximity (i.e.
R\Ss{O}/R\Ss{T}$>>$ 1) due to low gravity, debris, or an irregular
asteroid shape. In addition, the multiple stages of approach to the target 
make the overall orbital profiles to spread over a wide range of distances to the target.

\begin{figure*} \begin{center}
\includegraphics[width=5.2in]{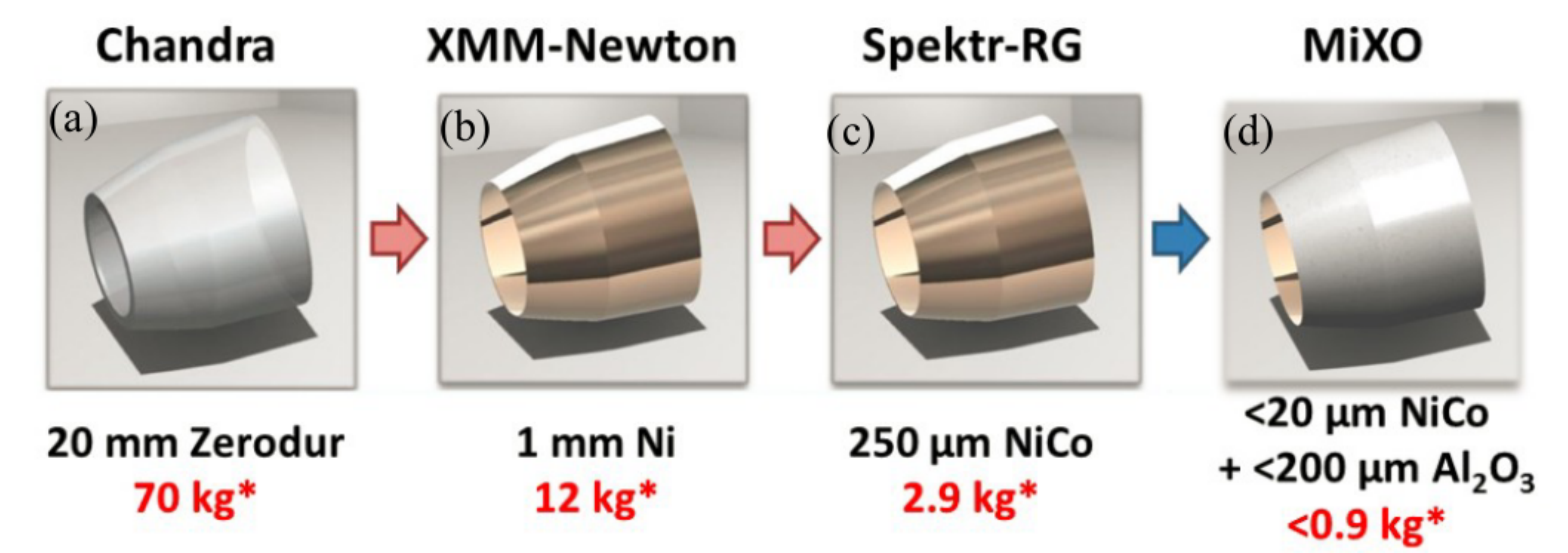} \\
\caption{ Advances of X-ray optics. (a) A ground and polished glass
substrate, (b) electroformed Ni shell, (c) thin electroformed NiCo
shell, and (d) a metal-ceramic hybrid shell. The weights (*) are for a
70 cm dia., 60 cm long mirror shell. For small mirrors ($\sim$5 cm dia.)
suitable for planetary science, a single shell would weigh $\sim$10 g or less.  }
\label{f:optics}
\end{center}
\end{figure*}

Imaging by
focusing X-ray optics enables a more forgiving orbit configuration
for observation, and focusing telescopes can continue to
accumulate useful data throughout the mission (R\Ss{T}/R\Ss{O} $\lesssim$ 300
for global measurements, R\Ss{T}/R\Ss{O} $\lesssim$ 100 for mapping).
As shown in Fig.~\ref{f:time}, the MiXO telescope would have completed
the same measurements as {\it NEAR-Shoemaker} or {\it Hayabusa} 
even before they reached ther observing distance. A
longer observing period enabled by MiXO also enhances a chance of success and new discoveries, since
detection of hard X-rays from heavy elements often relies on
relatively rare strong solar flares. In addition to the global measurement, with MiXO, the accumulated
data over $\sim$ 2 months (at 1 AU) during the quiet sun state alone
will be sufficient for detection of $\sim$1\% spatial variation for
major elements (e.g. O-K, Fe-L) over the entire surface.

The advantages of focusing optics extend to observations of
large planetary bodies as well.  For instance, at Mercury, the high thermal
loads force spacecraft to be at highly elliptical orbits,
limiting the observing duty cycle of collimator instruments. With focusing optics
less susceptible to changes in observing conditions,
the observing duty cycle can be maintained high.
In the case
of the Jovian system, high radiation environment (1 krad per an orbit,
depending on shielding, etc) severely limits the detector and 
shield size, and thus, the focusing optics is required to achieve large
light-collection power with a small detector.


\section{Advances of X-ray optics: Electroformed Nickel Replication with
Thermal Plasma Spray}
\label{s:enr}

While the scientific motivation and the advantages of X-ray
focusing optics have been known, it has been impractical to implement
conventional focusing X-ray telescopes in near-target planetary missions.
As aforementioned, typical planetary missions are a multi-instrument
discipline by nature, which is necessary 
but also limits the power, mass, volume and other resources allocated for
each instrument.  On the other hand, conventional X-ray optics tend to
be heavy and large.  This is because of
relatively low areal density (i.e.~relatively small collecting area per
unit mass) of X-ray mirrors. X-ray optics relies on grazing incidence for
efficient X-ray reflection, and a set of barrel shape mirrors are stacked
together in the Wolter-I configuration to establish needed effective area.
For instance, the outer X-ray mirror of {\it Chandra} is
about 1.4 m diameter by 1 m long with 10 m focal length, and weighs
about 1500 kg.

Fig.~\ref{f:optics} illustrates the advances of X-ray optics in weight
reduction over the years while maintaining high angular resolution
($<$15"--30") and improving collecting power.  In {\it Chandra}, high resolution mirrors were made
from directly grinding and polishing glass ceramic (zerodur), which
enabled sub arcsec angular resolution (Fig. \ref{f:optics}a). 
Electroformed Nickel Replication (ENR) process used to build X-ray mirrors
in {\it XMM-Newton} (Fig.~\ref{f:optics}b) utilizes relatively thin Ni
shell (1 mm, 8.9 g/cm\sS{3}) as a substrate
for the support of mirror figure and shape. The thin Ni shell enables reduction of the overall
mass by a factor of more than 5 compared to X-ray mirrors in {\it Chandra}
and allows larger light collecting area under the same packaging volume.
In ENR, the mirror figure and angular resolution is determined by the
surface quality of mandrel, and 10"--15" can be achieved routinely.
The Astronomical Roentgen Telescope (ART) \citep{Gubarev12} onboard
Spektr-RG, which is scheduled to launch in 2016, employs a NiCo alloy
(250 \um thick) as a
substrate instead of Ni under the ENR process, further reducing the
mirror mass (Fig.~\ref{f:optics}c).

\begin{figure} \begin{center}
\includegraphics[width=3.0in]{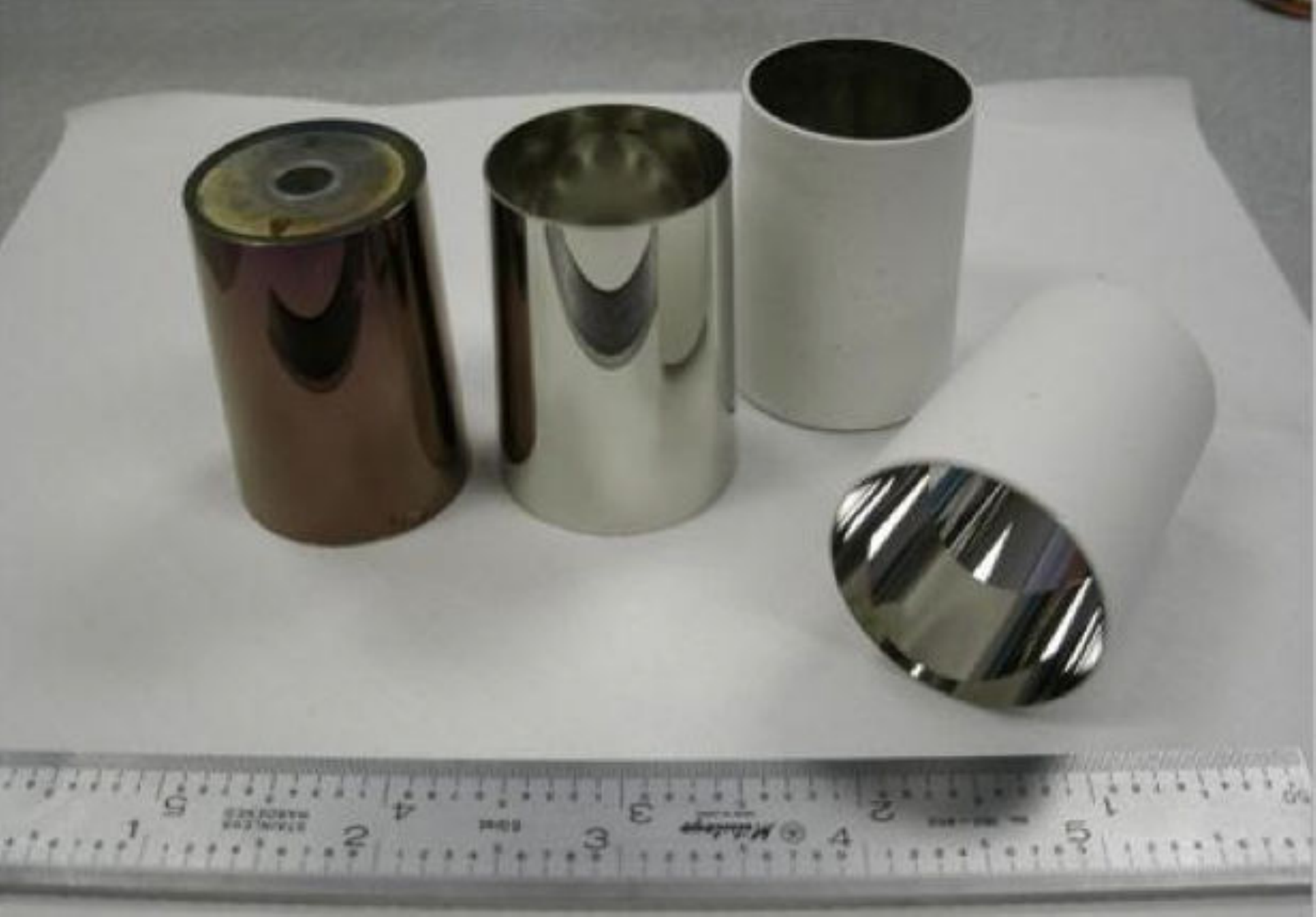} \\
\caption{A test single-bounce conical mandrel, a NiCo shell, 
two NiCo ceramic hybrid shells from left to right.
The width of the image spans about 14 cm.
}
\label{f:sample}
\end{center}
\end{figure}

Recently we have begun to develop a new approach, where we combine
the Plasma Thermal Spray (PTS) technology and the ENR process to form
metal-ceramic hybrid X-ray mirrors \citep{Romaine14}.  In the hybrid X-ray
mirrors, the lightweight ceramic layer provides the stiffness needed to
maintain the overall figure of X-ray optics, while the NiCo layer maintains the 
low surface roughness necessary for X-ray reflection.  By replacing the majority of
the NiCo layer ($\sim$8.9 g/cm\sS{3}) with
the lightweight alumina (Al\Ss{2}O\Ss{3}, 2.3--2.9 g/cm\sS{3}), we can
further reduce the mirror mass by another factor of 3 to 10 compared
to normal ENR mirrors, depending on the thickness of mirrors.
Fig.~\ref{f:sample} shows example modules: a conical mirror mandrel, a NiCo shell,
and two NiCo ceramic hybrid shells from left to right. The modules in
Fig.~\ref{f:sample} is single bounce, but ENR process can form a mirror with two
bounces in one package, simplifying the alignment structure without
introducing any degradation in angular resolution 
\citep[e.g.~see][]{Gubarev12, Romaine14}.
The hybrid shells in Fig.~\ref{f:sample}, which 
demonstrate a proof of concept, are made of an 300 \um
ceramic layer with 100 \um of a NiCo layer. Given the properties of NiCo
and ceramic layer, we expect that sub 20 \um NiCo + 200 \um ceramic X-ray
mirror shell can be achieved for relatively small shells (2 -- 20 cm
diameter).

\section{Example MiXO configurations and Performance}

MiXO made of metal-ceramic hybrid X-ray mirrors shows a tremendous promise in
enabling compact X-ray telescopes suitable for various types of planetary missions, which
can open a new era of sensitive X-ray imaging spectroscopy in planetary
science.  Until now, micro-pore optics (MPO) was considered the only
approach that can enable compact Wolter-I X-ray optics in 
planetary missions. The Mercury Imaging X-ray Spectrometer-T
(MIXS-T) on {\it Bepicolombo} employs the MPO and will be the first focusing X-ray telescope in
planetary science missions
\citep{Fraser10}.  In MPO of MIXS-T, the inner surface of small square pores or microchannels
($\sim$20 \um pitch) in a 0.9--2.2 mm thick plate is coated with
Ir for X-ray reflection. Then, mosaics of segmented MPO plates are assembled to
form a Wolter-I configuration. In this section, we estimate the performance
of MiXO using a few example configurations in comparison with {\it Bepicolombo}/MIXS-T.

\begin{figure} \begin{center}
\includegraphics[width=2.2in]{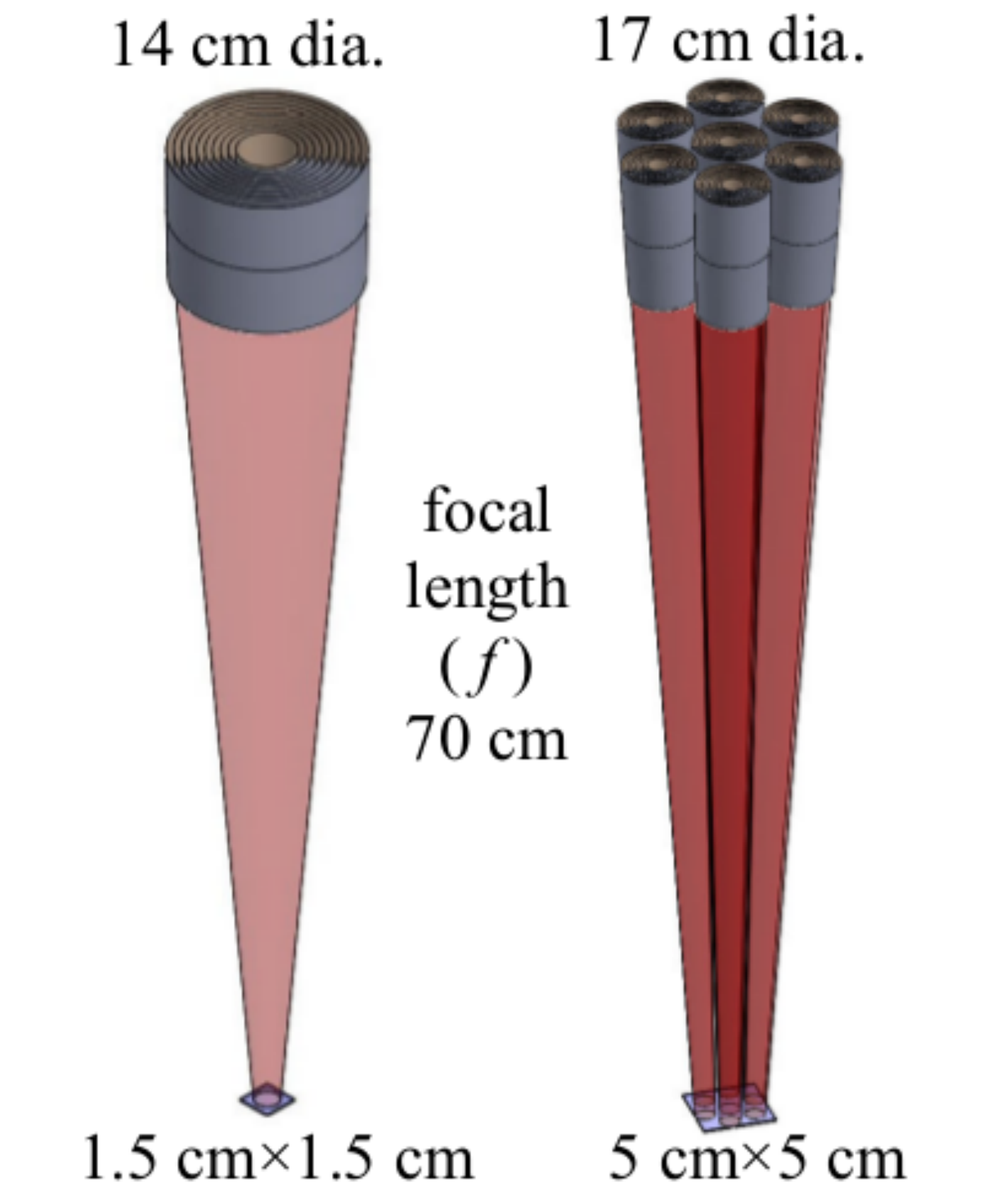} \\
\caption{Example MiXO telescopes:
(Left) the baseline design (MiXO-70B) and (Right) a wide-field option
(MiXO-70W). The red cones indicate
a 1.2 deg dia. FoV on the focal plane.
}
\label{f:ex}
\end{center}
\end{figure}

Fig.~\ref{f:ex} and Table \ref{t:ex} show two example configurations using the hybrid
shells at 70 cm focal length: the baseline option (MiXO-70B) (left)
and a wide field option (MiXO-70W) (right). Each shell in both the
options consists of 20 \um NiCo + 200 \um Al\Ss{2}O\Ss{3} layers. It is
$\sim$ 10 cm long ($\sim$ 5 cm for each bounce), and the gap between
the adjacent shells is 500 \um. The optics alone weighs
$\sim$ 0.9--1.5 kg, and the support structure will
add another 0.3--0.5 kg under a conservative estimate.
The baseline design is scalable to a smaller or larger
optics depending on available resources or mission
objectives. For example, a bigger telescope similar to
{\it Bepicolombo}/MIXS-T ($f$ $\sim$ 100 cm) or a smaller
telescope similar to JUXTA \citep[$f$ $\sim$ 25 cm][]{Ezoe13} can be
implemented.
MiXO-70W is optimized for
wide field monitoring with the combined FoV of 7 deg\sS2{} and broad
band coverage with relatively higher grasp at high energies due to
small grazing angles of each module.  The FoV of each module in
MiXO-70W is independent of each other, so that the alignment between
them is not critical.


\begin{table}
\caption{Example MiXO Telescopes}
\begin{tabular}{lll}
\hline\hline
Parameters	& MiXO-70B			& MiXO-70W 	\\
\hline
Focal Length ($f$)& 70 cm 			& 70 cm		\\
No. of Shells	& 50 				& 25 $\times$ 7	\\
Size (Diameter) & 4 -- 14 cm 			& 2 -- 5.6 cm \\
		&				& 17 cm overall \\
Mass (g) 	& 920 				& 210 $\times$ 7 	\\
Detector 	& 1.5 $\times$ 1.5 cm\sS{2}	& 5 $\times$ 5 cm\sS{2}	\\
FoV 		& 1 deg\sS{2} 			& 7 deg\sS{2}	\\
Ang. Resolution & 30$"$				& 30$"$	\\
Effective Area 	& 65 cm\sS{2} @ 1 keV		& 6.4 cm\sS{2} @ 1 keV \\
		& 15 cm\sS{2} @ 4 keV 		& 4.6 cm\sS{2} @ 4 keV \\
Grazing Angle	& 0.41 -- 1.43 deg		& 0.20 -- 0.51 deg  \\
\hline
\end{tabular}\\
*Assume 10\% reduction for alignment fixture
\label{t:ex}
\end{table}

\begin{figure} \begin{center}
\includegraphics[width=3.2in]{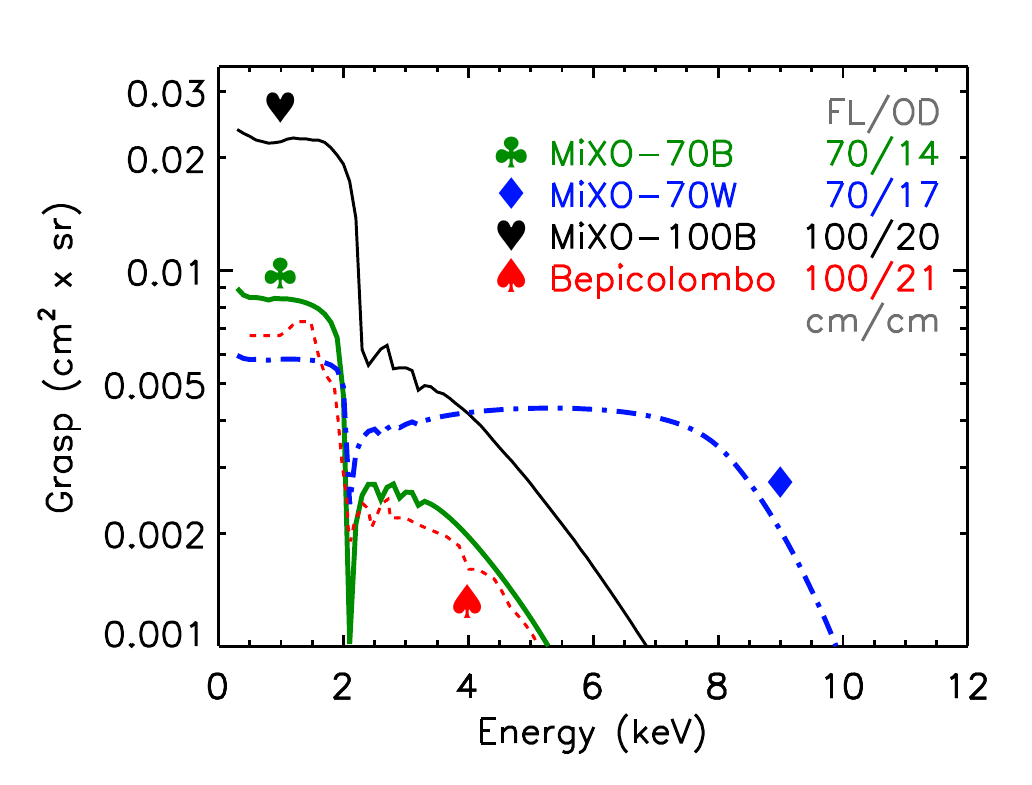} \\
\caption{Grasp of MiXO-70B, MiXO-70W and MiXO-100B optics with Ir
coating (10\% reduction for support structure, no surface or multilayer
optimization) in comparison with MIXS-T on {\it Bepicolombo}
\citep{Fraser10}. (FL: Focal Length, OD: Optics Diameter).
}
\label{f:perf}
\end{center}
\end{figure}

The sensitivity or light collection power of an X-ray telescope is
often described by the on-axis effective area.
Since X-ray emission from planetary bodies is often diffuse by nature especially 
when observed
near-by, another important parameter to describe the performance of X-ray telescopes
is grasp, which is the product of the effective area and the FoV.
Fig.~\ref{f:perf} compares the expected grasp of various MiXO optics with MIXS-T on {\it Bepicolombo}.
We assume that each shell of MiXO here is coated with Ir.
With only $\sim$1/3 of the packaging volume, MiXO-70B
({\color{dartmouthgreen} $\clubsuit$}) meets the performance of MIXS-T
({\color{red} $\spadesuit$}); or MiXO-100B ($\varheartsuit$ in
Fig.~\ref{f:perf}, the same baseline configuration as MiXO-70B but $f$
$\sim$100 cm with an optics envelope of 20 cm dia.), whose package is
similar to MIXS-T in size,
outperforms MIXS-T by a factor of $\sim$3.
In the case of MiXO-70W ({\color{blue} $\vardiamondsuit$}), small grazing angles boost high energy response,
enhancing a chance of the discovery of heavy elements.  Since our
approach enables additional optimizations such as polynomial profile of shell
geometry for uniform response over the wide field \citep{Conconi10} or multilayer coating
to enhance hard X-ray response \citep[e.g.~{\it NuSTAR};][]
{Harrison13}, one can further improve the response of MiXO telescopes.

In addition to outperforming on throughput (both on-axis effective area
and overall grasp) by a factor of 3 or more, our approach excels MPO in
two other areas: first, MPO can focus X-rays effectively only below 2 keV
due to the difficulty in coating the pore surface on the 20 \um scale.
For instance, multilayer coating cannot be applied to
small pores.
On the other hand, MiXO-70W in Fig.~\ref{f:perf} is designed for wide-field
hard ($>2$ keV) X-ray imaging and thus optimized for the detection and
localization of heavy elements such as rare-earth elements.

Second, the resolution of MPO is limited to 2' for lobster eye Angel
optics and 5--8' for Wolter-I optics. The additional degradation in
resolution for Wolter-I optics comes from misalignment among segmented
mirror facets in MPO. In fact, difficult alignment between segmented
mirrors requires a complex support structure, which
mostly erases its lightweight advantage relative to MiXO. 
On the other hand, in MiXO, each shell contains both bounces,
greatly simplifying alignment structure and mass, and a 15-30" resolution
(10--16 $\times$ improvement relative to MPO) can be achieved routinely.
Compared to MIXS-T, MiXO-70B, though a smaller telescope, improves the
detection sensitivity ($\propto$ PSF\_area\sS{1/2}) of small surface
features by $>$10$\times$  or extends the observing distance ($\propto$
PSF\_area\sS{1/4}) by $>$3$\times$ due
to relatively lower background ($\propto$ PSF\_area) of small PSFs
enabled by high resolution ($<$30").

High angular resolution planetary X-ray imaging 
with sub 100 cm\sS{2} effective area will be normally flux limited during
the quiet Sun phases.  On the other hand,
the full potential of the high angular resolution can be realized during
strong solar X-ray flares which can generate higher solar X-ray flux by a
few order of magnitudes.  Given the angular resolution in MiXO that is
comparable to NAVCAM on the Rosetta mission, a strong solar X-ray flare
can enable an X-ray snapshot of the detailed atomic composition, which
can be directly compared to the surface structure of comet nuclei and
outgassing jets in an optical image.

\section{Summary}

We have introduced a new approach to build lightweight focusing X-ray optics
with metal-ceramic hybrid X-ray mirrors through the ENR process and the
PTS technology. Hybrid X-ray mirrors are lighter by almost two orders
of magnitude relative to X-ray mirrors in {\it Chandra}. 
This enables compact lightweight X-ray telescopes which can be mounted
on resource-limited in-situ or near-target planetary missions.

We have illustrated the benefit of {\it true} imaging X-ray spectroscopy 
especially for observations of small planetary bodies such as asteroids
and comet nuclei.  With a few example configurations we have demonstrated
efficient light-collection power (3$\times$ improvement relative to an
alternative approach using MPO), high angular resolution ($>$10$\times$
improvement), high detection sensitivity ($>$10$\times$ improvement)
and wide energy band coverage (up to 10--15 keV with multilayer coating)
of MiXO. With MiXO, we can open a new era of planetary X-ray imaging
spectroscopy.

\section{Authors' Contribution}

JH carried out design and modeling work for MiXO, investigated its
feasibility and applications to various planetary missions, and
drafted the manuscript. SR demonstrated the proof of concept of MiXO
by fabricating and testing prototype MIXO modules.

\section{Acknowledgement}

The travel expense for the symposium was supported by NASA Grant
NNX12AG65G.  We thank Nicole D. Melso for performing simulations
to estimate the effective area and grasp for various telescope
configurations. We also thank Martin Elvis for the useful comments.



\begin{thebibliography}{}
%
%



\bibitem[Nittler et al.(2004)]{Nittler04}
Nittler LR et al. (2004) Bulk element compositions of meteorites: A
guide for interpreting remote-sensing geochemical measurements of
planets and asteroids. Antarctic Meteorite Research 17:231-251

\bibitem[Trombka et al.(2000)]{Trombka00}
Trombka JI et al. (2000) The Elemental Composition of Asteroid 433 Eros:
Results of the NEAR- Shoemaker X-ray Spectrometer. Science 289:2101-2105

\bibitem[Okada et al.(2006)]{Okada06}
Okada T et al. (2006) X-ray Fluorescence Spectrometry of Asteroid Itokawa
by Hayabusa. Science 312:1338-1341

\bibitem[Nittler et al.(2011)]{Nittler11}
Nittler LR et al. (2011) The Major-Element Composition of Mercury's
Surface from MESSENGER X-ray Spectrometry. Science 333:1847-1850

\bibitem[Binzel et al.(2010)]{Binzel10}
Binzel RP et al. (2010) Earth encounters as the origin of fresh surfaces
on near-Earth asteroids. Nature 463:331-334

\bibitem[Noguchi et al.(2011)]{Noguchi11}
Noguchi T et al. (2011) Incipient Space Weathering Observed on the
Surface of Itokawa Dust Particles. Science 333:1121-1125

\bibitem[Noguchi et al.(2014)]{Noguchi14}
Noguchi, T et al. (2014) Mineralogy of four Itokawa particles collected from the first touchdown site.
Earth, Planets, and Space  66:124-133 doi:10.1186/1880-5981-66-124

\bibitem[Altwegg et al.(2015)]{Altwegg15}
Altwegg K et al. (2015) 67P/Churyumov-Gerasimenko, a Jupiter family comet
with a high D/H ratio. Science doi:10.1126/science.1261952

\bibitem[Tachibana et al.(2013)]{Tachibana13}
Tachibana S et al. (2013) The Sampling System of Hayabusa-2: Improvements 
from the Hayabusa Sampler. Lunar and Planetary Science
Conference 44:1880-1881

\bibitem[Lauretta and O.-R. Team(2012)]{Lauretta12}
Lauretta D, O.-R. Team (2012) An Overview of the OSIRIS-REx Asteroid Sample Return
Mission. Lunar and Planetary Institute Science Conference 43:2491-2492

\bibitem[Hong and Romaine(2014)]{Hong14}
Hong J, Romaine S (2014) Miniature lightweight x-ray optics (MiXO) for
solar system exploration. SPIE vol 91441, p 91441F

\bibitem[Inamdar et al.(2014)]{Inamdar14}
Inamdar KN et al. (2014) Modeling the Expected Performance of the REgolith X-ray Imaging
Spectrometer (REXIS). SPIE vol 9222, p 922207

\bibitem[Gubarev et al.(2012)]{Gubarev12}
Gubarev M et al. (2012) The Marshall Space Flight Center development of mirror modules for
the ART-XC instrument aboard the Spectrum-Roentgen-Gamma mission. SPIE vol 8443, p 84431U 

\bibitem[Romaine et al.(2014)]{Romaine14}
Romaine S et al. (2014) Development of light weight replicated x-ray optics, II. SPIE vol
9144, p 91441H

\bibitem[Fraser et al.(2010)]{Fraser10}
Fraser GW et al. (2010) The mercury imaging X-ray spectrometer (MIXS) on bepicolombo. 
Planetary and Space Science 58:79-95

\bibitem[Ezoe et al.(2013)]{Ezoe13}
Ezoe Y et al. (2013) JUXTA: A new probe of X-ray emission from the Jupiter system. Advances
in Space Research  51:1605-1621

\bibitem[Conconi et al.(2010)]{Conconi10}
Conconi P et al. (2010) A wide field X-ray telescope for astronomical
survey purposes: from theory to practice.
Monthly Notices of the Royal Astronomical Society 405:877-886

\bibitem[Harrison et al.(2013)]{Harrison13}
Harrison F et al. (2013) The Nuclear Spectroscopic Telescope Array (NuSTAR) High-energy X-
Ray Mission. The Astrophysical Journal  770:103-121

\end{thebibliography}


\end{document}